\documentclass[twocolumn,amsmath,amssymb]{revtex4}
\usepackage{amsfonts}
\usepackage[usenames]{color}

\newcommand{\be}{\begin{equation}}
\newcommand{\ee}{\end{equation}}
\newcommand{\bea}{\begin{eqnarray}}
\newcommand{\eea}{\end{eqnarray}}

\begin{document}
%%%%%%%%%%%%%%%%%%%%%%%%%%%%%%%%%%%%%%%%%%%%%%%%%%%%%%%%%%%%%%%
\title{On gravity as a medium property in Maxwell equations}
\author{Jai-chan Hwang${}^{1}$, Hyerim Noh${}^{2}$}
\address{${}^{1}$Particle Theory  and Cosmology Group,
         Center for Theoretical Physics of the Universe,
         Institute for Basic Science (IBS), Daejeon, 34126, Republic of Korea
         \\
         ${}^{2}$Theoretical Astrophysics Group, Korea Astronomy and Space Science Institute, Daejeon, Republic of Korea
         }

%%%%%%%%%%%%%%%%%%%%%%%%%%%%%%%%%%%%%%%%%%%%%%%%%%%%%%%%%%%%%%%
%\date{\today}

%%%%%%%%%%%%%%%%%%%%%%%%%%%%%%%%%%%%%%%%%%%%%%%%%%%%%%%%%%%%%%%
\begin{abstract}

The effect of gravity in Maxwell's equations is often treated as a medium property. The commonly used formulation is based on managing Maxwell's equations in exactly the same form as in Minkowski spacetime and expressing the effect of gravity as a set of constitutive relations. We show that such a set of Maxwell's equations is, in fact, a combination of the electric and magnetic fields defined in two different non-covariant ways, both of which {\it fail} to identify the associated observer's four-vectors. The suggested constitutive relations are also {\it ad hoc} and unjustified. To an observer with a proper four-vector, the effect of gravity can be arranged as effective polarizations and magnetizations appearing in {\it both} the homogeneous and inhomogeneous parts. Modifying the homogeneous part by gravity is inevitable to any observer, and the result cannot be interpreted as the medium property. For optical properties one should directly handle Maxwell's equations in curved spacetime.

\end{abstract}
%%%%%%%%%%%%%%%%%%%%%%%%%%%%%%%%%%%%%%%%%%%%%%%%%%%%%%%%%%%%%%%

%%%%%%%%%%%%%%%%%%%%%%%%%%%%%%%%%%%%%%%%%%%%%%%%%%%%%%%%%%%%%%%
\maketitle

%%%%%%%%%%%%%%%%%%%%%%%%%%%%%%%%%%%%%%%%%%%%%%%%%%%%%%%%%%%%%%%
%
%
%
%%%%%%%%%%%%%%%%%%%%%%%%%%%%%%%%%%%%%%%%%%%%%%%%%%%%%%%%%%%%%%%
\section{Introduction}

Effect of gravity considered as a medium property in Maxwell's equations has a long history \cite{Eddington-1920, Whitehead-1922, Weyl-1923, Gordon-1923, Tamm-1924, Tamm-1925}. The result is often expressed as effective constitutive relations (permittivity, permeability, etc.) in the conventional Maxwell's equations in Minkowski spacetime \cite{Moller-1952, Plebanski-1960, deFelice-1971, Landau-Lifshitz-1971}. Recently, this was applied to the transformation optics where, using the behavior of photon in curved spacetime, metamaterials are used as a tool to inverse design the material to precisely achieve the desired photon path \cite{Leonhardt-2006}. The situation, however, is more complicated as the effect of gravity modifies Maxwell's equations even in the homogeneous part.

In special relativity it is well known that the electric and magnetic (EM) fields depends on the observer's motion through the observer's four-vector. This is related to the origin of special relativity \cite{Einstein-1905}. The situation can be handled using the observer's four-vector. In general relativity, the definition of EM fields depends on the metric through observer's four-vector which depends on the curved nature of spacetime \cite{Hwang-Noh-2023-EM-NL}. For an observer, the effect of gravity can be encoded in the effective polarizations and magnetizations appearing in both the homogeneous and inhomogeneous parts, thus as the medium properties of strange sorts demanding beyond the conventional constitutive relations of the material medium.

In \cite{Hwang-Noh-2023-EM-definition, Hwang-Noh-2023-EM-NL} we presented Maxwell's equations in curved spacetime using several different definitions of the EM fields. Two of the definitions are based on the covariant decomposition using the normal frame and the coordinate frame. The other two are based on non-covariant ways by simply identifying the EM fields as {\it components} of the field strength tensors, $F_{ab}$ and $F^*_{ab}$, respectively, with covariant indices in special relativistic forms. In all cases, the gravity effect is expressed using the polarization and magnetization vectors appearing in the homogeneous and/or inhomogeneous parts of Maxwell's equations.

However, the latter two non-covariant definitions are shown to be in trouble by lacking the corresponding four-vectors of the observer, thus failing to define the associated external charge and current densities even in the weak gravity limit \cite{Hwang-Noh-2023-EM-definition}. Without the four-vector identified such EM fields are not the ones which can be measured by any observer, thus disqualified as the EM fields. In \cite{Hwang-Noh-2023-EM-NL}, by considering a generic observer, we proved that, in gravitational environments, both the homogeneous and inhomogeneous parts are modified for any observer.

Previous literatures treating the gravity as a medium property in Maxwell's equations somehow rely on the latter two non-covariant definitions of the EM fields. Although the situation is in vacuum, the EM fields in the inhomogeneous and homogeneous equations are identified as ($D_i$, $H_i$) and ($E_i$, $B_i$), respectively, in non-covariant manner using components of $F^*_{ab}$ and $F_{ab}$, respectively. As mentioned, the variables defined in these two non-covariant ways are not the EM fields \cite{Hwang-Noh-2023-EM-NL}. We will show that the constitutive relations derived by connecting these two different definitions of the EM fields are {\it ad hoc} and unjustified.

Commonly used constitutive relations in the literature are the ones by Plebanski \cite{Plebanski-1960}. Plebanski, however, used the above mentioned two non-covariant definitions. M$\o$ller \cite{Moller-1952} (and Landau and Lifshitz \cite{Landau-Lifshitz-1971}) similarly introduced another non-covariant definitions of the EM fields now using a three-space metric in curved spacetime; whereas Plebanski used three-space Euclidean metric. We will show that these two methods lead to different Maxwell's equations, thus with different constitutive relations. Consequently medium properties are different depending on the non-covariant definitions.

In Sec.\ \ref{sec:normal} we begin with Maxwell's equations using the EM fields measured by an observer in the normal frame. In Sec.\ \ref{sec:medium} we manage the equations to identical forms (except for the charge and current densities) known in special relativity. The EM fields contrived in such a way will be shown to be the two non-covariant definitions mentioned above. Problems of the these definitions and the {\it ad hoc} nature of consequent constitutive relations are explained. Section \ref{sec:Discussion} is discussion. The covariant and two ($3+1$) decompositions are presented in Appendices \ref{sec:covariant} and \ref{sec:3+1}, respectively.

%%%%%%%%%%%%%%%%%%%%%%%%%%%%%%%%%%%%%%%%%%%%%%%%%%%%%%%%%%%%%%%
%
%
%
%%%%%%%%%%%%%%%%%%%%%%%%%%%%%%%%%%%%%%%%%%%%%%%%%%%%%%%%%%%%%%%
\section{Normal frame}
                                       \label{sec:normal}

The normal four-vector $n_a$ is normal to the hypersurface. It is the four-velocity of an observer instantaneously at rest in the chosen time slice, thus can be interpreted as an Eulerian observer as its motion follows the hypersurface independently of the coordinates chosen \cite{Smarr-York-1978, Wilson-Mathews-2003, Gourgoulhon-2012}. Therefore, independently of whether the effect of gravity can be treated as a medium property or not, Maxwell's equations in the normal frame is the one for proper optical analysis.

In the normal frame, Maxwell's equations become \cite{Hwang-Noh-2023-EM-NL}
\bea
   & & ( \sqrt{\overline h} \overline h{}^{ij} E_j )_{,i}
       = \sqrt{\overline h} \varrho,
   \label{Maxwell-normal-1} \\
   & &
       ( \sqrt{\overline h} \overline h{}^{ij} E_j )_{,0}
       - \eta^{ijk} \nabla_j
       ( N B_k - \eta_{k\ell m} \sqrt{\overline h} N^\ell
       \overline h{}^{mn} E_n )
   \nonumber \\
   & & \qquad
       = \sqrt{\overline h} N^i \varrho
       - {1 \over c} N \sqrt{\overline h} \overline h{}^{ij} j_j,
   \label{Maxwell-normal-2} \\
   & & ( \sqrt{\overline h} \overline h{}^{ij} B_j )_{,i} = 0,
   \label{Maxwell-normal-3} \\
   & & ( \sqrt{\overline h} \overline h{}^{ij} B_j )_{,0}
       + \eta^{ijk} \nabla_j ( N E_k
       + \eta_{k\ell m} \sqrt{\overline h} N^\ell
       \overline h{}^{mn} B_n )
   \nonumber \\
   & & \qquad
       = 0,
   \label{Maxwell-normal-4}
\eea
where indices of $E_i$, $B_i$, $j_i$, and $\eta_{ijk}$ are raised and lowered using $\delta_{ij}$ and its inverse, whereas $N$, $N_i$, and $\overline h_{ij}$ are the Arnowitt-Deser-Misner (ADM) metric variables introduced in Appendix \ref{sec:3+1} with the indices associated with the intrinsic metric $\overline h_{ij}$ and its inverse $\overline h{}^{ij}$; $\overline h \equiv {\rm det}(\overline h_{ij})$. For derivation, see Appendix \ref{sec:3+1}.

These can be arranged as
\bea
   & & ( E^i + P_{\rm E}^i )_{,i}
       = \sqrt{\overline h} \varrho,
   \label{Maxwell-normal-PM-1} \\
   & &
       ( E^i + P_{\rm E}^i )_{,0}
       - \eta^{ijk} \nabla_j
       ( B_k - M^{\rm E}_k )
   \nonumber \\
   & & \qquad
       = \sqrt{\overline h} N^i \varrho
       - {1 \over c} N \sqrt{\overline h} \overline h{}^{ij} j_j,
   \label{Maxwell-normal-PM-2} \\
   & & ( B^i + P_{\rm B}^i )_{,i} = 0,
   \label{Maxwell-normal-PM-3} \\
   & & ( B^i + P_{\rm B}^i )_{,0}
       + \eta^{ijk} \nabla_j ( E_k - M^{\rm B}_k ) = 0,
   \label{Maxwell-normal-PM-4}
\eea
where the effective polarization and magnetization vectors, ${\bf P}$s and ${\bf M}$s, caused by the metric are
\bea
   & & P_{\rm E}^i
       \equiv \sqrt{\overline h} \overline h{}^{ij} E_j - E^i,
   \nonumber \\
   & & M^{\rm E}_i
       \equiv (1 - N) B_i
       + \eta_{ijk} \sqrt{\overline h} N^j \overline h{}^{k\ell} E_\ell,
   \nonumber \\
   & & P_{\rm B}^i
       \equiv \sqrt{\overline h} \overline h{}^{ij} B_j - B^i,
   \nonumber \\
   & & M^{\rm B}_i
       \equiv (1 - N) E_i
       - \eta_{ijk} \sqrt{\overline h} N^j \overline h{}^{k\ell} B_\ell.
   \label{PM-normal}
\eea
While the structure of Maxwell's equations is preserved, the effect of gravity appears as effective ${\bf P}$s and ${\bf M}$s in {\it both} the homogeneous and inhomogeneous parts.

%%%%%%%%%%%%%%%%%%%%%%%%%%%%%%%%%%%%%%%%%%%%%%%%%%%%%%%%%%%%%%%
%
%
%
%%%%%%%%%%%%%%%%%%%%%%%%%%%%%%%%%%%%%%%%%%%%%%%%%%%%%%%%%%%%%%%
\section{Gravitation as a medium property}
                                       \label{sec:medium}

%%%%%%%%%%%%%%%%%%%%%%%%%%%%%%%%%%%%%%%%%%%%%%%%%%%%%%%%%%%%%%%
\subsection{By new field redefinitions}

Here we derive the conventionally used constitutive relations by transformation of the normal frame EM fields. Introducing new EM variables
\bea
   & & \hskip -.8cm
       \hat E_i \equiv N E_i
       + \eta_{ijk} \sqrt{\overline h} N^j
       \overline h{}^{k\ell} B_\ell, \quad
       \hat B^i \equiv \sqrt{\overline h}
       \overline h{}^{ij} B_j,
   \nonumber \\
   & & \hskip -.8cm
       {\breve E}^i \equiv \sqrt{\overline h}
       \overline h{}^{ij} E_j, \quad
       {\breve B}_i \equiv N B_i
       - \eta_{ijk} \sqrt{\overline h} N^j
       \overline h{}^{k\ell} E_\ell,
   \label{EB-relation-normal}
\eea
Maxwell's equations can be written in a simple form as
\bea
   & & {\breve E}^i_{\;\;,i}
       = \sqrt{\overline h} \varrho,
   \label{Maxwell-hat-breve-1} \\
   & & {\breve E}^i_{\;\;,0}
       - \eta^{ijk} \nabla_j {\breve B}_k
       = \sqrt{\overline h} N^i \varrho
       - {1 \over c} N \sqrt{\overline h} \overline h{}^{ij} j_j,
   \label{Maxwell-hat-breve-2} \\
   & & {\hat B}^i_{\;\;,i} = 0,
   \label{Maxwell-hat-breve-3} \\
   & & {\hat B}^i_{\;\;,0}
       + \eta^{ijk} \nabla_j {\hat E}_k = 0,
   \label{Maxwell-hat-breve-4}
\eea
which are, in the absence of the charge and current densities, formally the same as Maxwell's equations in special relativity. Still, notice the two different notations for the EM fields, ($\hat E_i$, $\hat B_i$) versus ($\breve E_i$, $\breve B_i$), and yet another notation for the charge and current densities ($\varrho$, $j_i$) associated withe the normal frame. The meaning of ($\hat E_i$, $\hat B_i$) and ($\breve E_i$, $\breve B_i$) as independent definitions of the EM fields will become clear later, see Eqs.\ (\ref{Fab-hat}) and (\ref{Fab-breve}).

In \cite{Plebanski-1960, deFelice-1971, Leonhardt-2006}, $\breve E_i$ and $\breve B_i$ are {\it identified} as $\hat {D}_i$ and $\hat {H}_i$, respectively, as if these are associated with $\hat E_i$ and $\hat B_i$ in a medium. Thus, we have
\bea
   & & \hat D^i_{\;\;,i} = {1 \over c} \sqrt{-g} J^0,
   \label{Maxwell-P-1} \\
   & & \hat D^i_{\;\;,0} - \eta^{ijk} \nabla_j \hat H_k
       = - {1 \over c} \sqrt{-g} J^i,
   \label{Maxwell-P-2} \\
   & & \hat B^i_{\;\;,i} = 0,
   \label{Maxwell-P-3} \\
   & & \hat B^i_{\;\;,0} + \eta^{ijk} \nabla_j \hat E_k = 0,
   \label{Maxwell-P-4}
\eea
where we used the four-current.

The effective constitutive relations are introduced as
\bea
   \hat {D}^i
       = \varepsilon^{ij} \hat E_j
       + \gamma^{ij} \hat {H}_j, \quad
       \hat B^i = \mu^{ij} \hat {H}_j
       - \gamma^{ij} \hat E_j,
\eea
where $\varepsilon^{ij}$ and $\mu^{ij}$ are the permittivity and permeability tensors, respectively, and $\gamma^{ij}$ is the electromagnetic mixing term; indices of $\varepsilon_{ij}$, etc., are associated with $\delta_{ij}$ and its inverse. From Eq.\ (\ref{EB-relation-normal}), or using Eq.\ (36) in \cite{Hwang-Noh-2023-EM-NL}, we can show
\bea
   & & \varepsilon^{ij} = \mu^{ij}
       = {N \sqrt{\overline h} \over N^2 - N^k N_k}
       \left( \overline h^{ij} - {N^i N^j \over N^2} \right)
       = {\sqrt{-g} \over - g_{00}} g^{ij},
   \nonumber \\
   & &
       \gamma^{ij} = {\eta^{ijk} N_k \over
       N^2 - N^\ell N_\ell}
       = \eta^{ijk} {g_{0k} \over - g_{00}}.
   \label{constitutive}
\eea
Equations (\ref{Maxwell-P-1})-(\ref{constitutive}) were derived in \cite{Plebanski-1960}, see Sec.\ \ref{sec:P} below, and by a transformation from the coordinate frame in \cite{deFelice-1971}; our derivation above is a transformation from the normal frame. As long as the final aim is to present Maxwell's equations in exactly special relativistic form, the results are the same and unique.

In the weak gravity limit, we have
\bea
   g_{00} = - (1 + 2 \alpha), \quad
       g_{0i} = 0, \quad
       g_{ij} = (1 + 2 \varphi) \delta_{ij}.
   \label{WG-metric}
\eea
The metric perturbations are related to the gravitational potential as $\alpha \equiv \Phi/c^2$ and $\varphi \equiv - \Psi/c^2$ with $\Psi = \Phi$ in many situations \cite{Noh-Hwang-Bucher-2019}. For a spherically symmetric system with a mass $M$ at the center, we have $\varphi = - \alpha = GM/(rc^2)$. Our metric variables become
\bea
    & & N = 1 + \alpha, \quad
        N_i = 0 = N^i, \quad
        \overline h_{ij} = (1 + 2 \varphi) \delta_{ij},
    \nonumber \\
    & &
        \overline h^{ij} = (1 - 2 \varphi) \delta^{ij}, \quad
        \overline h = 1 + 6 \varphi.
\eea
Equation (\ref{constitutive}) gives
\bea
   \varepsilon^{ij} = \mu^{ij}
       = (1 + \varphi - \alpha) \delta^{ij}, \quad
       \nu^{ij} = 0,
\eea
thus $\mu = \varepsilon = 1 + \varphi - \alpha$ with $\varepsilon_{ij} \equiv \varepsilon \delta_{ij}$ and $\mu_{ij} \equiv \mu \delta_{ij}$, and
\bea
   n = {c \over v} = \sqrt{\varepsilon \mu}
       = 1 + \varphi - \alpha
       = 1 - 2 \alpha
       = 1 + 2 {GM \over r c^2}.
   \label{n-P}
\eea
This gives the correct value of refractive index $n$ in the gravitational field \cite{Schneider-1992, Petters-2002, Meneghetti-2021}. Assuming $\mu = \varepsilon = {\rm constant}$, and ignoring the four-current, the same result also follows directly from Maxwell's equations in Eqs.\ (\ref{Maxwell-P-1})-(\ref{Maxwell-P-4}) and in Eqs.\ (\ref{Maxwell-normal-1})-(\ref{Maxwell-normal-4}). Thus, the gravity-medium analogy based on the {\it wrong} equations happens to give the same value of refractive index as the one based on Maxwell's equations in the normal frame.

In the remainder of this section, we explain why Eqs.\ (\ref{Maxwell-P-1})-(\ref{Maxwell-P-4}) and the constitutive relations in Eq.\ (\ref{constitutive}) are {\it wrong}.

%%%%%%%%%%%%%%%%%%%%%%%%%%%%%%%%%%%%%%%%%%%%%%%%%%%%%%%%%%%%%%%
\subsection{Nature of the new field variables}

Based on the formal similarity of Eqs.\ (\ref{Maxwell-P-1})-(\ref{Maxwell-P-4}) with Maxwell's equations in Minkowski spacetime, in \cite{Plebanski-1960, deFelice-1971, Leonhardt-2006} the effect of gravitation (encoded in the metric variables $N$, $N_i$, and $\overline h_{ij}$) is interpreted as effective constitutive relations of an optical medium. Such a gravity-medium correspondence argument based on a specially arranged form of Maxwell's equations in Eqs.\ (\ref{Maxwell-hat-breve-1})-(\ref{Maxwell-hat-breve-4}) is in error as the EM fields as well as charge and current densities depend on the observer, whereas no such observers can be attached to the EM variables used in these equations.

The errors are in (i) $\hat E_i$ and $\hat B_i$ are defined in a non-covariant way, (ii) $\breve E_i$ and $\breve B_i$ are defined in a yet another non-covariant way, (iii) both of these two definitions fail to identify the observers who could measure such EM fields (without the associated observer, such variables are {\it not} EM fields by any sense), and consequently, (iv) one cannot properly introduce the corresponding charge and current densities which should be measured by the same observers, and (v) the identification of ($\breve E_i$, $\breve B_i$) with ($\hat {D}_i$, $\hat {H}_i$) and consequent constitutive relations are {\it ad hoc} and unjustified.

As mentioned, Eqs.\ (\ref{Maxwell-hat-breve-1})-(\ref{Maxwell-hat-breve-4}) are written using variables from three different definitions, for two of which (i and ii) one cannot identify the associated observers who can measure such EM fields. The normal frame, on the other hand, is associated with the Eulerian observer who is at rest and normal to the hypersurface \cite{Smarr-York-1978, Wilson-Mathews-2003, Gourgoulhon-2012}.

We can understand errors in (i)-(iv) by showing how we can derive Eqs.\ (\ref{Maxwell-hat-breve-1})-(\ref{Maxwell-hat-breve-4}) directly. This will reveal the meaning of the EM variables used in the equations. As in Minkowski spacetime, by na\"ively assigning the EM fields as tensor components of $F_{ab}$ in a non-covariant way as
\bea
   F_{0i} \equiv - {\hat E}_i, \quad
       F_{ij} \equiv \eta_{ijk} \hat B{}^k,
   \label{Fab-hat}
\eea
the second one in Eq.\ (\ref{Maxwell-tensor-eqs}) gives Eqs.\ (\ref{Maxwell-hat-breve-3}) and (\ref{Maxwell-hat-breve-4});
however, the first one gives quite complicated equations; see Eqs.\ (54) and (55) in \cite{Hwang-Noh-2023-EM-NL}.

In this way of assigning physical variables to {\it components} of a tensor in a special form, in curved spacetime, one cannot find the four-vector of an observer who can measure the EM fields defined in Eq.\ (\ref{Fab-hat}); see also \cite{Crater-1994}. This is true even in the weak gravity limit \cite{Hwang-Noh-2023-EM-definition}; this is because, in general, the six conditions used in Eq.\ (\ref{Fab-hat}) cannot be satisfied by using only three conditions available for the normalized time-like four-vector, which is true even considering the gauge conditions available in relativistic  perturbation theory, see \cite{Hwang-Noh-2023-EM-NL} and below Eq.\ (\ref{EM-potenial-normal}). In \cite{Hwang-Noh-2023-EM-NL}, we proved that gravity causes modification in the homogeneous and inhomogeneous parts inevitably for any observer. Without the four-vector of the observer, one cannot introduce the corresponding charge and current densities which are defined using the same four-vector. Such a combination of variables cannot be considered as the EM fields as these combinations cannot be measured by any observer.

The same argument applies to Eqs.\ (\ref{Maxwell-hat-breve-1}) and (\ref{Maxwell-hat-breve-2}). By defining the EM fields in another non-covariant way as
\bea
   F{}^*_{0i} \equiv - {\breve B}_i, \quad
       F{}^*_{ij}
       \equiv - \eta_{ijk} \breve E{}^k,
   \label{Fab-breve}
\eea
the first one in Eq.\ (\ref{Maxwell-tensor*-eqs}) leads to
\bea
   & & {\breve E}^i_{\;\;,i}
       = {1 \over c} \sqrt{-g} J^0,
   \label{Maxwell-mixed-1} \\
   & & {\breve E}^i_{\;\;,0}
       - \eta^{ijk} \nabla_j {\breve B}_k
       = - {1 \over c} \sqrt{-g} J^i,
   \label{Maxwell-mixed-2}
\eea
while the second one gives quite complicated equations; see Eqs.\ (60) and (61) in \cite{Hwang-Noh-2023-EM-NL}. We kept the components of the four-current as we cannot identify the observer's four-vector. In terms of the charge and current densities in the normal frame, these become Eqs.\ (\ref{Maxwell-hat-breve-1}) and (\ref{Maxwell-hat-breve-2}).
%The second one in Eq.\ (\ref{Maxwell-tensor*-eqs}) gives a complicated counterpart which is ignored; see Eqs.\ (43) and (44) in \cite{Hwang-Noh-2023-EM-NL}.
Setting $\sqrt{-g} F^{ab}$ in special relativistic form, as often used in \cite{Plebanski-1960, deFelice-1971, Leonhardt-2006}, is the same as setting $F^*_{ab}$ in special relativistic form, see Eq.\ (\ref{F*ab-Plebanski}).

%%%%%%%%%%%%%%%%%%%%%%%%%%%%%%%%%%%%%%%%%%%%%%%%%%%%%%%%%%%%%%%
\subsection{A popular misconception}

In the literature of gravitational wave detection using electromagnetic methods, only Eq.\ (\ref{Fab-hat}) is adopted without referring to Eq.\ (\ref{Fab-breve}), see \cite{Berlin-2022, Domcke-2022}.
One motivation is based on a misconception, confusing $F_{ab}$ and the homogeneous Maxwell's equations expressed in terms of the four-potential as the evidence for no metric interference in both, explicitly stated recently in \cite{Domcke-2022}. If we further adopt Eq.\ (\ref{Fab-breve}), using ($\hat D_i$, $\hat H_i$) instead of ($\breve E_i$, $\breve B_i$) as we did above Eq.\ (\ref{Maxwell-P-1}), we recover the gravity-medium analogy with the constitutive relations in Eq.\ (\ref{constitutive}). Thus, the gravity medium analogy literatures share the same misconception we clarify below.

It is well known that using the four-potential introduced as $F_{ab} \equiv A_{b;a} - A_{a;b}$ the metric disappears, as the connections cancel away, thus
\bea
   F_{ab} = A_{b,a} - A_{a,b},
   \label{four-potential}
\eea
even in curved Riemannian spacetime. As a consequence, the homogeneous Maxwell's equations in the second of Eq.\ (\ref{Maxwell-tensor-eqs}) is identically satisfied, and in this way, only the inhomogeneous Maxwell's equations ``exhibits a gravitational influence," see Chapter X of \cite{Whitehead-1922}. Here, one should not confuse the implication. This well known fact is true only in terms of the four-potential. In terms of the EM fields, presence of gravity on the homogeneous Maxwell's equations is unavoidable, and metric appears in any decomposition of $F_{ab}$ into the EM fields using the observer's four-vector. There exists no conflict because relations between the EM field and the four-potential are nontrivial in curved spacetime, see \cite{Thorne-MacDonald-1982}. Let us elaborate this point.

From Eq.\ (\ref{EB-def}), in the normal frame, we have $E_i = F_{ib} n^b$ and $B_i \equiv F^*_{ib} n^b$. Using Eqs.\ (\ref{four-potential}) and (\ref{n_a-ADM}), we can derive
\bea
   & & E_i = {1 \over N} \left[ \partial_i A_0
       - A_{i,0}
       - ( \partial_i A_j - \partial_j A_i ) N^j \right],
   \nonumber \\
   & & B_i
       = {1 \over \sqrt{\overline h}}
       \overline h_{ij} \eta^{jk\ell} \partial_k A_\ell.
   \label{EM-potenial-normal}
\eea
Thus, metric appears in the relations, and this is true for any observer's four-vector \cite{Hwang-Noh-2023-EM-NL}; synchronous gauge conditions with $N = 1$ and $N_i = 0$ can make $E_i$ simple, but the metric still remains in $B_i$. The EM fields simply related to the four-potential in special relativistic form are nothing other than $\hat E_i$ and $\hat B_i$, i.e.,
\bea
   \hat E_i = \partial_i A_0 - \partial_0 A_i, \quad
       \hat B_i = \eta_{ijk} \partial^j A^k,
   \label{EM-potenial}
\eea
which has no physical meaning in curved spacetime as we explained above and in \cite{Hwang-Noh-2023-EM-definition, Hwang-Noh-2023-EM-NL}. Therefore, using the EM fields, presence of metric in $F_{ab}$ and in the homogeneous Maxwell's equations are inevitable in curved spacetime.

%%%%%%%%%%%%%%%%%%%%%%%%%%%%%%%%%%%%%%%%%%%%%%%%%%%%%%%%%%%%%%%
\subsection{Plebanski}
                                          \label{sec:P}

Most of the literature concerning gravity-medium analogy follow Plebanski \cite{Plebanski-1960}. Without paying attention to the nature of the EM fields defined in two different ways, using the notation in a medium, Plebanski {\it defined} the EM fields in vacuum as
\bea
   & & F_{0i} \equiv - \hat E_i, \quad
       F_{ij} \equiv \eta_{ijk} \hat B^k,
   \nonumber \\
   & & \sqrt{-g} F^{0i} \equiv \hat D^i, \quad
       \sqrt{-g} F^{ij} \equiv \eta^{ijk} \hat H_k.
\eea
The second line is the same as
\bea
   & & F^*_{0i} \equiv - \hat H_i, \quad
       F^*_{ij} \equiv - \eta_{ijk} \hat D^k.
   \label{F*ab-Plebanski}
\eea
From Eq.\ (\ref{Maxwell-tensor-eqs}), with $H^{ab}$ replaced by $F^{ab}$, we have Eqs.\ (\ref{Maxwell-P-1})-(\ref{Maxwell-P-4}). Equation (\ref{constitutive}) can be derived from
\bea
   \sqrt{-g} F^{ab} = \sqrt{-g} g^{ac} g^{bd} F_{cd}.
   \label{H-F}
\eea

These definitions are the same as Eqs.\ (\ref{Fab-hat}) and (\ref{Fab-breve}). Thus, as explained, in this non-covariant way of defining the EM fields, one cannot identify the four-vector allowing these definitions and consequently cannot introduce the accompanying charge and current densities, even to the linear order in gravity \cite{Hwang-Noh-2023-EM-definition}.

The error (v), stating the {\it ad hoc} and unjustified nature of the constitutive relations derived based on defining the EM fields in the two non-covariant ways, can be shown using an example of similarly derived constitutive relations of \cite{Moller-1952, Landau-Lifshitz-1971} which differs from the one in \cite{Plebanski-1960, deFelice-1971, Leonhardt-2006}. We present the alternative analysis of \cite{Moller-1952, Landau-Lifshitz-1971} in the following.

%%%%%%%%%%%%%%%%%%%%%%%%%%%%%%%%%%%%%%%%%%%%%%%%%%%%%%%%%%%%%%%
\subsection{M$\o$ller and Landau-Lifshitz}
                                          \label{sec:LL}

In a problem of Sec.\ 90 in \cite{Landau-Lifshitz-1971}, Landau and Lifshitz {\it defined} the EM fields as
\bea
   & & F_{0i} \equiv - \overline E_i, \quad
       F_{ij} \equiv \overline \eta_{ijk} \overline B{}^k,
   \nonumber \\
   & &
       \sqrt{-g_{00}} F^{0i} \equiv \overline D{}^i, \quad
       \sqrt{-g_{00}} F^{ij} \equiv \overline
       \eta{}^{ijk} \overline H_k,
   \label{EBDH-LL}
\eea
where indices of $\overline \eta_{ijk}$ and $\overline E_i$ etc. in this section are associated with $\overline \gamma_{ij}$ in Eq.\ (\ref{metric-LL}). The same definitions can be found in Sec.\ 115 of M$\o$ller \cite{Moller-1952}; we note that M$\o$ller's work is earlier than Plebanski and Landau and Lifshitz (including all editions). In terms of $F^*_{ab}$ the definition in the second line becomes
\bea
   & & F^*_{0i} = - \overline H_i, \quad
       F^*_{ij} = - \overline \eta_{ijk} \overline D{}^k.
   \label{Fab*-LL}
\eea
Thus, $\overline E_i$ and $\overline B_i$ are defined assuming $F_{ab}$ is in special relativistic form, and $\overline D_i$ and $\overline H_i$ are defined assuming $F^*_{ab}$ is in special relativistic form, where spatial indices are associated with $\overline \gamma_{ij}$ as the metric. Notice that in these definitions metric still appears in definitions of $\overline B{}^k$ and $\overline H{}^k$ as we have $\overline \eta_{ijk} = \sqrt{\overline \gamma} \eta_{ijk}$ and $\overline B{}^i \equiv \overline \gamma{}^{ij} \overline B_j$, etc.

As in \cite{Plebanski-1960}, the EM fields are introduced in non-covariant manner by assigning the physical fields to components of tensors in certain forms, and we {\it cannot} identify the four-vectors of the observers who can measure such fields. That is, in curved spacetime, we can show that the definitions in the first line of Eq.\ (\ref{EBDH-LL}) are not possible for any choice of the four-vector (observer), and the same is true for definitions in Eq.\ (\ref{Fab*-LL}).

Maxwell's equations in Eq.\ (\ref{Maxwell-tensor-eqs}) give
\bea
   & & {1 \over \sqrt{\overline \gamma}}
       \left( \sqrt{\overline \gamma} \overline D{}^i
       \right)_{,i}
       = {1 \over c} \overline N J^0,
   \label{Maxwell-LL-1} \\
   & & {1 \over \sqrt{\overline \gamma}}
       \left( \sqrt{\overline \gamma} \overline D{}^i
       \right)_{,0}
       - \overline \eta{}^{ijk} \nabla_j \overline H_k
       = - {1 \over c} \overline N J^i,
   \label{Maxwell-LL-2} \\
   & & {1 \over \sqrt{\overline \gamma}}
       \left( \sqrt{\overline \gamma} \overline B{}^i
       \right)_{,i}
       = 0,
   \label{Maxwell-LL-3} \\
   & & {1 \over \sqrt{\overline \gamma}}
       \left( \sqrt{\overline \gamma} \overline B{}^i
       \right)_{,0}
       + \overline \eta{}^{ijk} \nabla_j \overline E_k
       = 0,
   \label{Maxwell-LL-4}
\eea
where we kept the four-current as we {\it cannot} identify the observer's four-vector associated with the definitions of EM fields in Eq.\ (\ref{EBDH-LL}).

In Eq.\ (\ref{EBDH-LL}) the EM fields are related as
\bea
   \hskip -.2cm
   \overline D{}^i = {1 \over \overline N} \overline E{}^i
       - \overline \eta{}^{ijk} {\overline N_j
       \over \overline N{}^2} \overline H_k, \quad
       \overline B{}^i = {1 \over \overline N} \overline H{}^i
       + \overline \eta{}^{ijk} {\overline N_j
       \over \overline N{}^2} \overline E_k,
   \label{relations-LL}
\eea
which follows from Eq.\ (\ref{H-F}). Introducing the effective constitutive relations as
\bea
   \overline D{}^i
       = \overline \varepsilon{}^{ij} \overline E_j
       + \overline \nu{}^{ij} \overline H_j, \quad
       \overline B^i = \overline \mu{}^{ij} \overline H_j
       - \overline \nu{}^{ij} \overline E_j,
\eea
where indices of $\overline \varepsilon^{ij}$, etc., are associated with $\overline \gamma_{ij}$, we have
\bea
   \overline \varepsilon{}^{ij} = \overline \mu{}^{ij}
       = {\overline \gamma{}^{ij} \over \overline N}, \quad
       \overline \nu{}^{ij} = \overline \eta{}^{ijk}
       {\overline N_k \over \overline N{}^2},
   \label{constitutive-LL-overline}
\eea
thus, the permittivity and permeability become $\varepsilon = \mu = 1/\overline N$. These results were presented in Sec.\ 115 of \cite{Moller-1952} and Sec.\ 90 of \cite{Landau-Lifshitz-1971}.

In order to compare with Plebanski's constitutive relations in Eq.\ (\ref{constitutive}), we present results using the EM field vectors associated with $\delta_{ij}$ as the metric. We introduce $\overline B_i \equiv B_i$ where the index of $B_i$ is associated with $\delta_{ij}$, and similarly for $E_i$, $D_i$, and $H_i$. From Eq.\ (\ref{EBDH-LL}), we can show
\bea
   & & F_{0i} = - E_i, \quad
       F_{ij} = \sqrt{\overline \gamma} \eta_{ijk}
       \overline \gamma{}^{k\ell} B_\ell,
   \nonumber \\
   & & \sqrt{-g} F^{0i} = \sqrt{\overline \gamma}
       \overline \gamma{}^{ij} D_j, \quad
       \sqrt{-g} F^{ij} = \eta^{ijk} H_k.
   \label{EBDH-LL-2}
\eea
The second-line can be written as
\bea
   & & F^*_{0i} = - H_i, \quad
       F^*_{ij} = - \sqrt{\overline \gamma} \eta_{ijk}
       \overline \gamma{}^{k\ell} D_\ell.
   \label{Fab*-LL-2}
\eea
In this form, nontrivial metric dependence appears explicitly. Maxwell's equations become
\bea
   & & {1 \over \sqrt{\overline \gamma}}
       \left( \sqrt{\overline \gamma} \overline \gamma{}^{ij}
       D_j \right)_{,i} = {1 \over c} \overline N J^0,
   \label{Maxwell-LL2-1} \\
   & & {1 \over \sqrt{\overline \gamma}}
       \left( \sqrt{\overline \gamma} \overline \gamma{}^{ij}
       D_j \right)_{,0}
       - {1 \over \sqrt{\overline \gamma}}
       \eta^{ijk} \nabla_j H_k
       = - {1 \over c} \overline N J^i,
   \label{Maxwell-LL2-2} \\
   & & {1 \over \sqrt{\overline \gamma}}
       \left( \sqrt{\overline \gamma} \overline \gamma{}^{ij}
       B_j \right)_{,i} = 0,
   \label{Maxwell-LL2-3} \\
   & & {1 \over \sqrt{\overline \gamma}}
       \left( \sqrt{\overline \gamma} \overline \gamma{}^{ij}
       B_j \right)_{,0}
       + {1 \over \sqrt{\overline \gamma}}
       \eta^{ijk} \nabla_j E_k
       = 0.
   \label{Maxwell-LL2-4}
\eea
From Eq.\ (\ref{relations-LL}) we have relations
\bea
   & & D_i = {1 \over \overline N} E_i
       - \sqrt{\overline \gamma} \eta_{ijk}
       {\overline N{}^j \over \overline N{}^2}
       \overline \gamma{}^{k\ell} H_\ell,
   \nonumber \\
   & & B_i = {1 \over \overline N} H_i
       + \sqrt{\overline \gamma} \eta_{ijk}
       {\overline N{}^j \over \overline N{}^2}
       \overline \gamma{}^{k\ell} E_\ell,
   \label{relations-LL-2}
\eea
and the constitutive relations
\bea
   \mu^j_i = \varepsilon^j_i
       = {1 \over \overline N} \delta^j_i, \quad
       \nu^j_i = - \sqrt{\overline \gamma} \eta_{ik\ell}
       {\overline N{}^k \over \overline N{}^2}
       \overline \gamma{}^{\ell j},
\eea
where indices of $\mu_{ij}$ etc., are associated with $\delta_{ij}$. Thus, we have $\mu = \varepsilon = 1/\overline N$. In terms of the spacetime metric, we have constitutive relations
\bea
   \varepsilon^j_i = \mu^j_i
       = {1 \over \sqrt{-g_{00}}} \delta^j_i, \quad
       \nu^j_i = - \eta_{ik\ell} {\sqrt{-g}
       \over \sqrt{-g_{00}}} g^{0k} g^{j\ell},
   \label{constitutive-LL}
\eea
which {\it differs} from Eq.\ (\ref{constitutive}).

In the weak gravity limit in Eq.\ (\ref{WG-metric}), both Eq.\ (\ref{constitutive-LL-overline}) and Eq.\ (\ref{constitutive-LL}) give $\overline \nu{}^{ij} = 0 = \nu^j_i$ and
\bea
   \mu = \varepsilon = 1 - \alpha, \quad
       \sqrt{\mu \varepsilon} = 1 - \alpha
       = 1 + {GM \over rc^2},
   \label{n-LL}
\eea
which misses a $2$-factor compared with Eq.\ (\ref{n-P}). As we show below, here we {\it cannot} identify the medium property $\sqrt{\mu\varepsilon}$ as the refractive index $n \equiv c/v$ with $v$ the phase velocity. Directly from Maxwell's equations in Eqs.\ (\ref{Maxwell-LL2-1})-(\ref{Maxwell-LL2-4}), we have
\bea
   & & [ (1 + \varphi) \varepsilon E^i ]_{,i} = 0,
   \\
   & & [ (1 + \varphi) \varepsilon E^i ]_{,0}
       = \eta^{ijk} \nabla_j ( \mu^{-1} B_k ),
   \\
   & & [ (1 + \varphi) B^i ]_{,i} = 0,
   \\
   & & [ (1 + \varphi) B^i ]_{,0}
       = - \eta^{ijk} \nabla_j E_k.
\eea
{\it Assuming} $\alpha$ and $\varphi$ constant in time and space, we have
\bea
   B^i_{\;\;,00} = (1 - 2 \varphi) {1 \over \mu \varepsilon}
       \Delta B^i,
\eea
and similarly for $E^i$. Thus, we can identify
\bea
   n \equiv {c \over v} = (1 + \varphi)
       \sqrt{\mu \varepsilon} = 1 - \alpha + \varphi
       = 1 + 2 {GM \over r c^2},
\eea
and recover the correct index of refraction in the gravitational field.

Not only the two constitutive relations in Eqs.\ (\ref{constitutive}) and (\ref{constitutive-LL}) differ from each other, in \cite{Moller-1952, Landau-Lifshitz-1971} the homogeneous part of Maxwell's equations in Eqs.\ (\ref{Maxwell-LL2-3}) and (\ref{Maxwell-LL2-4}) are also modified which cannot be interpreted as the medium property. In \cite{Landau-Lifshitz-1971} the authors mention referring to special relativistic analogy of Eqs.\ (\ref{Maxwell-LL-1})-(\ref{Maxwell-LL-4}) that ``the analogy (purely formal, of course)," suggesting their definition only as an exercise problem.

Using the non-covariant definitions of EM fields in gravity systems similarly as in this section, the dependence of medium property on the spatial decomposition method and the coordinate used is clarified in \cite{Gibbons-Werner-2019}.

%%%%%%%%%%%%%%%%%%%%%%%%%%%%%%%%%%%%%%%%%%%%%%%%%%%%%%%%%%%%%%%
%
%
%
%%%%%%%%%%%%%%%%%%%%%%%%%%%%%%%%%%%%%%%%%%%%%%%%%%%%%%%%%%%%%%%
\section{Discussion}
                                   \label{sec:Discussion}

In order to treat the effect of gravity in Maxwell's equations as the property of an optical medium, in the previous literature \cite{Moller-1952, Plebanski-1960, deFelice-1971, Landau-Lifshitz-1971, Leonhardt-2006}, the analyses are often made based on Eqs.\ (\ref{Maxwell-P-1})-(\ref{Maxwell-P-4}) which, for vanishing charge and current densities, look identical to the ones in Minkowski spacetime, with effective constitutive relations in Eq.\ (\ref{constitutive}).

Here we showed that these are {\it not} a consistent set of Maxwell's equations. Instead, these are a combination of {\it two} sets of Maxwell's equations with only the simpler parts written; the omitted complementary parts are highly complicated, see \cite{Hwang-Noh-2023-EM-NL}. Furthermore, due to the absence of the associate four-vectors for both definitions, we cannot identify the associated observers and consequently cannot properly introduce the charge and current densities for any of these EM fields. In \cite{Hwang-Noh-2023-EM-NL} we showed that no observer can measure such EM fields defined in non-covariant manner. Thus, these are {\it not} EM fields, and even the individual homogeneous and inhomogeneous parts are {\it not} parts of proper Maxwell's equations.

The association of the EM fields defined in two different ways as if these are related in a medium, with consequent effective constitutive relations in Eq.\ (\ref{constitutive}), is also {\it ad hoc} and unwarranted. This is shown by the two different constitutive relations derived from the two similarly introduced non-covariant definitions in \cite{Plebanski-1960, deFelice-1971, Leonhardt-2006} and \cite{Moller-1952, Landau-Lifshitz-1971}.

Historically, the simple nature of Eqs.\ (\ref{Maxwell-hat-breve-1})-(\ref{Maxwell-hat-breve-4}) has guided Einstein to propose the four-dimensional Maxwell's equations, previously constructed by Minkowski in special relativity \cite{Minkowski-1907}, now valid in general curved spacetime \cite{Einstein-1916}. Although used the two separate coincidences with special relativity as the guide, Einstein has not proposed neither of these as the definition of the EM fields. He has clearly noticed the nature of two different definitions of the EM fields and the complicated relationship among them.

As the normal frame is associated with the Eulerian observer, Eqs.\ (\ref{Maxwell-normal-1})-(\ref{Maxwell-normal-4}) is the proper Maxwell's equations to study the optical property measured by an Eulerian observer. These are arranged as Eqs.\ (\ref{Maxwell-normal-PM-1})-(\ref{Maxwell-normal-PM-4}) so that the effect of gravitation behave as effective polarization and magnetization vectors, ${\bf P}$s and ${\bf M}$s, appearing in {\it both} the homogeneous as well as inhomogeneous Maxwell's equations. The ${\bf P}$ and ${\bf M}$ present in the homogeneous Maxwell's equations cannot be interpreted as a medium property.

The ${\bf P}$ and ${\bf M}$ in electromagnetism is an alternative way of presenting the constitutive relations \cite{Born-Wolf-2019}. The optical property of light depends on medium property represented by the ${\bf P}$ and ${\bf M}$ and is governed by Maxwell's equations \cite{Born-Wolf-2019}. Similarly, in gravitational environments the optical property is determined by Maxwell's equations and depends on the effective ${\bf P}$s and ${\bf M}$s caused by gravity which depends on the frame choice. We propose the normal frame as the one to use because it is associated with an Eulerian observer \cite{Smarr-York-1978, Wilson-Mathews-2003, Gourgoulhon-2012}. %How to handle the gravity caused effective ${\bf P}$s and ${\bf M}$s as the optical medium property is open for future study.

%\vskip .5cm
%%%%%%%%%%%%%%%%%%%%%%%%%%%%%%%%%%%%%%%%%%%%%%%%%%%%%%%%%%%%%%%
%
%
%
%%%%%%%%%%%%%%%%%%%%%%%%%%%%%%%%%%%%%%%%%%%%%%%%%%%%%%%%%%%%%%%
%\begin{center}
%{\bf Declaration of competing interest}
%\end{center}

%The authors declare that they have no known competing financial interests or personal relationships that could have appeared to influence the work reported in this paper.

%%%%%%%%%%%%%%%%%%%%%%%%%%%%%%%%%%%%%%%%%%%%%%%%%%%%%%%%%%%%%%%
%
%
%
%%%%%%%%%%%%%%%%%%%%%%%%%%%%%%%%%%%%%%%%%%%%%%%%%%%%%%%%%%%%%%%
\section*{Acknowledgments}

We thank Professor Gary Gibbons for clarifying comments. H.\ N.\ was supported by the National Research Foundation (NRF) of Korea funded by the Korean Government (No.\ 2018R1A2B6002466 and No.\ 2021R1F1A1045515). J.\ H.\ was supported by IBS under the project code, IBS-R018-D1, and by the NRF of Korea funded by the Korean Government (No.\ NRF-2019R1A2C1003031).

\appendix
%%%%%%%%%%%%%%%%%%%%%%%%%%%%%%%%%%%%%%%%%%%%%%%%%%%%%%%%%%%%%%%
%
%
%
%%%%%%%%%%%%%%%%%%%%%%%%%%%%%%%%%%%%%%%%%%%%%%%%%%%%%%%%%%%%%%%
\section{Covariant formulation}
                                      \label{sec:covariant}

Maxwell's equations in a medium using the electromagnetic field strength tensors are \cite{Minkowski-1907}
\bea
   H^{ab}_{\;\;\;\;;b}
       = {1 \over c} J^a, \quad
       \eta^{abcd} F_{bc,d} = 0,
   \label{Maxwell-tensor-eqs}
\eea
where $H_{ab}$ is associated with ($D_a$, $H_a$) as $F_{ab}$ is associated with ($E_a$, $B_a$). Using dual tensors, $F^*_{ab} \equiv {1 \over 2} \eta_{abcd} F^{cd}$ and similarly for $H^*_{ab}$, we have
\bea
   - {1 \over 2} \eta^{abcd} H^*_{bc,d}
       = {1 \over c} J^a,
       \quad
       F^{*ab}_{\;\;\;\;\;\;;b} = 0.
   \label{Maxwell-tensor*-eqs}
\eea
Notice the ordinary derivatives appearing in the second of Eq.\ (\ref{Maxwell-tensor-eqs}) and the first of Eq.\ (\ref{Maxwell-tensor*-eqs}). We have
\bea
   \eta_{0ijk} = - \sqrt{-g} \eta_{ijk}, \quad
       \eta^{0ijk} = {1 \over \sqrt{-g}} \eta^{ijk},
   \label{eta}
\eea
where $g \equiv {\rm det}(g_{ij})$ and indices of $\eta_{ijk}$ are raised and lowered using $\delta_{ij}$ and its inverse. In vacuum, $H_{ab} = F_{ab}$.

Using a generic time-like four-vector $u_a$ of an observer, with $u_a u^a \equiv -1$, we define the EM fields, $E_a$ and $B_a$, associated with the four-vector as \cite{Moller-1952, Lichnerowicz-1967, Ellis-1973}
\bea
   & & F_{ab} \equiv u_a E_b - u_b E_a - \eta_{abcd} u^c B^d,
   \nonumber \\
   & & F^*_{ab} = u_a B_b - u_b B_a + \eta_{abcd} u^c E^d,
   \label{Fab-cov}
\eea
with $E_a u^a \equiv 0 \equiv B_a u^a$. The EM fields are
\bea
   E_a = F_{ab} u^b, \quad
       B_a = F_{ab}^* u^b.
   \label{EB-def}
\eea
The current four-vector is decomposed using the {\it same} four-vector as
\bea
   J^a \equiv \varrho c u^a + j^a, \quad
       j_a u^a \equiv 0,
   \label{J^a}
\eea
where $\varrho$ and $j_a$ are charge and current densities, respectively, measured by the same observer. The covariant form of Maxwell's equations in terms of $E_a$ and $B_a$ are derived in \cite{Ellis-1973}, see also \cite{Hwang-Noh-2022-Axion-EM}. In other sections we often use tildes to indicate the spacetime covariant quantities.

%%%%%%%%%%%%%%%%%%%%%%%%%%%%%%%%%%%%%%%%%%%%%%%%%%%%%%%%%%%%%%%
%
%
%
%%%%%%%%%%%%%%%%%%%%%%%%%%%%%%%%%%%%%%%%%%%%%%%%%%%%%%%%%%%%%%%
\section{Two ($3+1$) decompositions}
                             \label{sec:3+1}

We introduce two different decompositions of the spacetime metric. The ADM metric and its inverse are \cite{ADM}
\bea
   & & \hskip -.8cm
       g_{00} \equiv - N^2 + N^i N_i, \quad
       g_{0i} \equiv N_i, \quad
       g_{ij} \equiv \overline h_{ij},
   \nonumber \\
   & & \hskip -.8cm
       g^{00} = - {1 \over N^2}, \quad
       g^{0i} = {N^i \over N^2}, \quad
       g^{ij} = \overline h{}^{ij}
       - {N^i N^j \over N^2},
   \label{ADM-metric-def}
\eea
where $\overline h{}^{ij}$ is an inverse of the three-space intrinsic metric $\overline h_{ij}$, thus $\overline h{}^{ik} \overline h_{jk} \equiv \delta^i_j$, and the index of $N_i$ is raised and lowered by $\overline h_{ij}$ and its inverse. We have
\bea
   \eta_{0ijk} = - N \overline \eta_{ijk}, \quad
       \eta{}^{0ijk} = {1 \over N}
       \overline \eta{}^{ijk},
\eea
where indices of $\overline \eta_{ijk}$ is raised and lowered using $\overline h_{ij}$ and its inverses.

The normal frame is introduced as
\bea
   n_i \equiv 0, \quad
       n_0 = - N, \quad
       n{}^i = - {N^i \over N}, \quad
       n{}^0 = {1 \over N}.
   \label{n_a-ADM}
\eea
For the EM fields and current density in the normal frame, we set
\bea
   \widetilde B_i \equiv \overline B_i, \quad
       \widetilde B_0 = \overline B_i N^i, \quad
       \widetilde B^i = \overline B{}^i, \quad
       \widetilde B^0 = 0,
\eea
and similarly for $\overline E_a$, and $\overline j_a$ with $\widetilde E_i \equiv \overline E_i$ etc.; indices of $\overline B_i$ etc. are raised and lowered by $\overline h_{ij}$ and its inverse. Based on the normal frame, Eqs.\ (\ref{Fab-cov}) and (\ref{J^a}) give
\bea
   & & F_{0i} = - N \overline E_i
       - \overline \eta_{ijk} N^j \overline B{}^k, \quad
       F_{ij} = \overline \eta_{ijk} \overline B{}^k,
   \nonumber \\
   & & F{}^*_{0i} = - N \overline B_i
       + \overline \eta_{ijk} N^j \overline E{}^k, \quad
       F{}^*_{ij} = - \overline \eta_{ijk}
       \overline E{}^k,
   \nonumber \\
   & & J^0 = \varrho c {1 \over N},
       \quad
       J^i = - \varrho c {N^i \over N}
       + \overline j^i.
   \label{Fab-ADM}
\eea
By introducing $\overline B_i \equiv B_i$ with the index of $B_i$ raised and lowered using $\delta_{ij}$ and its inverse, and similarly for $E_i$ and $j_i$, from Eq.\ (\ref{Maxwell-tensor-eqs}) we can derive Eqs.\ (\ref{Maxwell-normal-1})-(\ref{Maxwell-normal-4}), see also \cite{Hwang-Noh-2023-EM-NL}.

In \cite{Moller-1952, Landau-Lifshitz-1971} an alternative ($3+1$) decomposition is used with (in our notation)
\bea
   & & g_{00} \equiv - \overline N{}^2, \quad
       g_{0i} \equiv \overline N_i, \quad
       g_{ij} = \overline \gamma_{ij}
       - {\overline N_i \overline N_j
       \over \overline N{}^2},
   \label{metric-LL} \\
   & & g^{00} = - {1 \over \overline N{}^2}
       \left( 1 - {\overline N{}^k \overline N_k \over
       \overline N{}^2} \right), \quad
       g^{0i} = {\overline N{}^i \over \overline N{}^2}, \quad
       g^{ij} \equiv \overline \gamma{}^{ij},
   \nonumber
\eea
where the index of $\overline N_i$ is associated with $\overline \gamma_{ij}$ and its inverse metric $\overline \gamma^{ij}$; in \cite{Landau-Lifshitz-1971}, $\overline N{}^2 \equiv h$ and $\overline N_i \equiv h g_i$. We have
\bea
   & & \eta_{0ijk} = - \sqrt{-g} \eta_{ijk}
       = - \overline N \overline \eta_{ijk}
       = - \sqrt{-g_{00}} \overline \eta_{ijk},
   \nonumber \\
   & & \eta^{0ijk} = {1 \over \sqrt{-g}} \eta^{ijk}
       = {1 \over \overline N} \overline \eta^{ijk}
       = {1 \over \sqrt{-g_{00}}} \overline \eta^{ijk},
   \label{eta-LL}
\eea
where $\sqrt{-g} = \overline N \sqrt{\overline \gamma} = N \sqrt{\overline h}$ with $\overline \gamma \equiv {\rm det}(\overline \gamma_{ij})$; here, indices of $\overline \eta_{ijk}$ is associated with $\overline \gamma_{ij}$ as the metric. Compared with the ADM decomposition, we have
\bea
   & & ds^2 = - \overline N{}^2 \left( d x^0
       - {\overline N_i \over \overline N{}^2} d x^i \right)
        \left( d x^0
       - {\overline N_j \over \overline N{}^2} d x^j \right)
   \nonumber \\
   & & \qquad
       + \overline \gamma_{ij} d x^i d x^j,
   \\
   & & d s^2 = - N^2 d x^0 d x^0
   \nonumber \\
   & & \qquad
       + \overline h_{ij} \left( d x^i + N^i d x^0 \right)
       \left( d x^j + N^j d x^0 \right).
\eea

We can express the normal four-vector, the EM fields, field strength tensor and four-current in the normal frame using this decomposition. But these are more complicated compared with Eqs.\ (\ref{n_a-ADM})-(\ref{Fab-ADM}) for the ADM decomposition. For example, we have
\bea
   & & \hskip -.8cm
       n_i \equiv 0, \quad
       n_0 = - {\overline N^2 \over
       \sqrt{ \overline N^2 - \overline N^k \overline N_k}},
   \nonumber \\
   & & \hskip -.8cm
       n{}^i = - {\overline N^i \over
       \sqrt{ \overline N^2 - \overline N^k \overline N_k}}, \quad
       n{}^0 = {\sqrt{ \overline N^2 - \overline N^k \overline N_k} \over \overline N^2}.
\eea
This decomposition is more adapted to the coordinate frame with
\bea
   \bar n_i = {\overline N_i \over \overline N}, \quad
       \bar n_0 = - \overline N, \quad
       \bar n{}^i \equiv 0, \quad
       \bar n{}^0 = {1 \over \overline N},
\eea
etc., but Maxwell's equations are still more complicated compared with the compact forms available in the normal frame in the ADM decomposition, see Eqs.\ (20)-(23) in \cite{Hwang-Noh-2023-EM-NL}.

%%%%%%%%%%%%%%%%%%%%%%%%%%%%%%%%%%%%%%%%%%%%%%%%%%%%%%%%%%%%%%%%%
%
%   References
%
%%%%%%%%%%%%%%%%%%%%%%%%%%%%%%%%%%%%%%%%%%%%%%%%%%%%%%%%%%%%%%%%%

%%%%%%%%%%%%%%%%%%%%%%%%%%%%%%%%%%%%%%%%%%%%%%%%%%%%%%%%%%%%%%%

%%%%%%%%%%%%%%%%%%%%%%%%%%%%%%%%%%%%%%%%%%%%%%%%%%%%%%%%%%%%%%%
\end{document}